# Electrochemically induced hyperfluorescence based on the formation of charge-transfer excimers

Chang-Ki Moon[1,2], Yuka Yasuda[3], Yu Kusakabe[3], Anna Popczyk[1], Shohei Fukushima[3], Julian F. Butscher[1], Nachiket Pathak[1], Hironori Kaji[3]*, Malte C. Gather[1,2]*

[1]Humboldt Centre for Nano- and Biophotonics, Institute for Light and Matter, Department of Chemistry and Biochemistry, University of Cologne, Greinstr. 4-6, 50939 Köln, Germany

[2]Organic Semiconductor Centre, School of Physics and Astronomy, University of St Andrews, North Haugh, St Andrews KY16 9SS, United Kingdom

[3]Institute for Chemical Research, Kyoto University, Uji, Kyoto, Japan

Keywords: Electrochemiluminescence, hyperfluorescence, electrochemically induced hyperfluorescence, charge-transfer excimer, energy transfer, spectroelectrochemistry, calligraphic display

Used extensively in sensing applications, the application of solution-state electrochemiluminescent devices (ECLDs) in lighting and displays has been constrained by their low luminance and short operational lifetime. Here, we report a record improvement in the luminance, efficiency, and operational longevity of ECLDs by introducing electrochemically induced hyperfluorescence (ECiHF) and demonstrate its use in a calligraphic display. We use the "double-decker" arrangement assumed by the electron donor and acceptor segments of the molecule TpAT-tFFO to realize thermally activated delayed fluorescence from an electrogenerated charge-transfer (CT) excimer state and a subsequent energy transfer to the rubrene emitter TBRb. ECLDs based on this strategy achieve an unprecedented luminance of >6,200 cd/m² and their operational lifetime is more than 10-fold longer than all previous ECLDs with meaningful efficiency or brightness. We identify energy level alignment between excimer and emitter as a crucial factor for efficient ECiHF; spectroelectrochemical analysis reveals that devices with energy gaps < 0.4 eV operate on a pure excimer mechanism across a wide range of frequencies. Our findings highlight the potential of ECiHF for improving ECLDs and pave the way to commercial applications of this form of "fluid light".

By combining electrochemistry with light emission, electrochemiluminescence (ECL) has become a versatile tool in biomedical research, particularly for sensitive detection of biomolecules in immunoassays and diagnostics[1-3], food safety analysis[4], and environmental surveillance[5]. Unlike other types of electrically generated light emission that rely on charge carrier transport and recombination in solid state, such as organic light-emitting diodes (OLEDs)[6] and light-emitting electrochemical cells[7,8], ECL generally occurs by ionic reactions in liquid or gel state, thus providing great flexibility in device form factor and greatly simplifying fabrication[9-12]. To extend the ECL technology to future applications in lighting and display applications, and hence to further benefit from the cost-effectiveness and scalability of this "fluid light", recent research has explored novel luminophores, electrode configurations, and operating mechanisms to enhance the brightness, efficiency, and operational stability of electrochemiluminescent devices (ECLDs)[13-18]. Achieving intense ECL generally necessitates production of abundant radical ions and their subsequent rapid radiative recombination. However, the accumulation of radical ions can trigger side reactions that deteriorate the luminophores in the device[19], limiting the maximum brightness of ECLDs that rely on this "annihilation process" to less than 700 cd/m$^2$ and restricting their operational times to a few minutes at most.

In OLEDs, luminophores supporting thermally activated delayed fluorescence (TADF) have gained great popularity as they enhance the fraction of emissive singlet excitons to ~100% by rapidly depopulating non-emissive triplet states via reverse intersystem crossing (RISC)[20,21]. In ECLDs, TADF emitters have also been demonstrated to achieve up to a 4-fold improvement in ECL efficiency[22-24]; however, so far with limited benefit to the maximum achievable luminance and operational stability. Using TADF emitters in ECLDs also presents additional challenges, particularly related to solvent selection, as positive solvatochromism leads to significant spectral shifts and broadening of the emission in high-polarity solvents[25,26].

Hyperfluorescence (HF), which combines a conventional fluorescent emitter with a triplet sensitizer[27], offers a promising alternative that enables high luminance and efficiency without introducing substantial spectral shifts. In our previous exciplex-based ECLDs[28,29], the exciplex state served as a triplet sensitizer[30,31]; however, the RISC process on the exciplex was not sufficiently fast to support efficient HF.

In this paper, we realize electrochemically induced hyperfluorescence (ECiHF) through the electrogeneration of charge-transfer (CT) excimers—rather than exciplexes—, thus achieving rapid RISC and subsequent energy transfer to a fluorescent emitter. Our study investigates two TADF molecules, the recently reported TpAT-tFFO[32] and TpATtBu-tFFO, a previously unreported analogue; both feature face-to-face alignment of the donor and acceptor components, thus exhibiting excellent TADF characteristics. Using concentration-dependent photoluminescence (PL) spectroscopy in solution, we observe a transition from monomer emission to excimer emission while preserving clear hallmarks of TADF as the concentration of TpAT-tFFO increases. Adding a rubrene dye to these high-concentration TpAT-tFFO solutions then leads to the appearance of HF and enables ECiHF in AC-driven ECLDs, resulting in record-high luminance, efficiency, and operational stability. To showcase the benefits of this enhanced light emission, we demonstrate an ECL display featuring a calligraphic electrode design. Finally, to gain insights into molecular design rules for future high-performance ECLDs, we compare the excimers of TpAT-tFFO and TpATtBu-tFFO. The latter has a higher energy and shows lower performance in ECLDs with a stronger dependence on operating voltage and frequency. Absorption spectroelectrochemistry confirms excimer formation as the primary operating mechanism in TpAT-tFFO devices, and reveals a mixed operating mechanism for TpATtBu-tFFO devices, in which excimer formation and ionic annihilation compete.

A schematic illustration of the ECiHF mechanism is given in Fig. 1a. By alternately between positive and negative voltages at the electrodes, oxidized and reduced forms of the TADF molecules accumulate near each electrode surface. These redox products recombine to electrostatically bound CT excimers, with the highest occupied molecular orbital (HOMO) and lowest unoccupied molecular orbital (LUMO) localized on donor and acceptor segments of the TADF molecules, respectively. This spatial separation results in minimal orbital overlap and thus enables efficient intermolecular CT transitions. The CT excimers undergo intersystem crossing (ISC) and RISC cycles, and transfer singlet energy to neighboring fluorescent emitters via FRET, thereby achieving an exciton utilization exceeding the 25% spin-statistical limit.

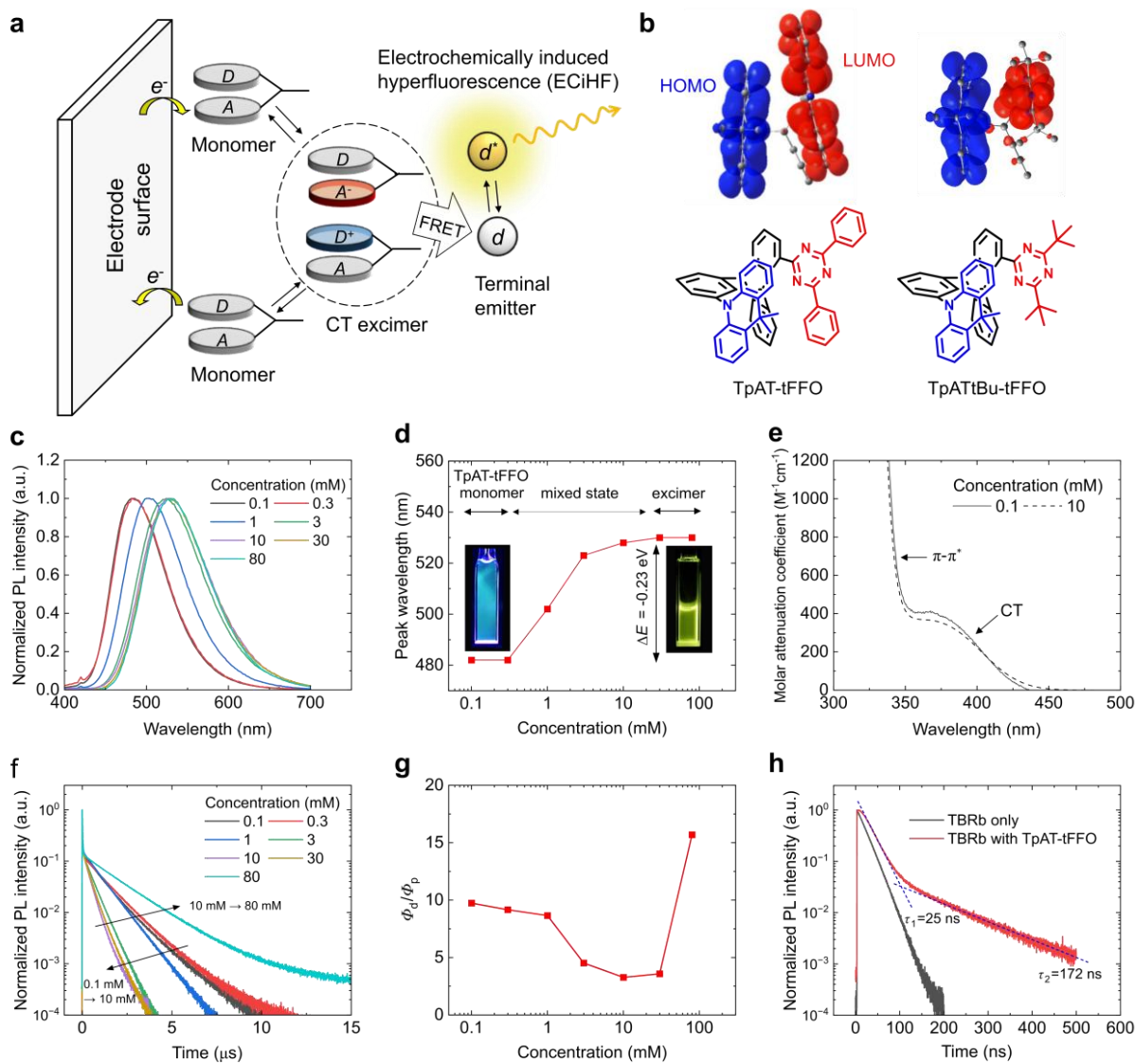

**Fig. 1. Electrochemically induced hyperfluorescence (ECiHF). a**, Schematic of ECiHF under AC operation with formation of CT excimers and energy transfer to the terminal emitter. **b**, Molecular structures of TpAT-tFFO and TpATtBu-tFFO in three-dimensional configurations, with the donor and acceptor groups represented as blue and red segments, respectively, in the line drawings. For both molecules, the HOMO (blue surface) and LUMO (red surface) states are separated on the donor and acceptor groups, respectively. **c**, Photoluminescence (PL) spectra of TpAT-tFFO in a 2:1 by volume mixture of toluene and acetonitrile. **d**, Peak wavelengths of PL spectra for TpAT-tFFO at concentrations ranging from 0.1 mM to 80 mM. **e**, Molar attenuation of TpAT-tFFO measured at concentrations of 0.1 mM and 10 mM. **f**, Transient PL of TpAT-tFFO at concentrations ranging from 0.1 mM to 80 mM. **g**, Ratio of the quantum yield of delayed to prompt fluorescence ($\Phi_d/\Phi_p$) against TpAT-tFFO concentration. **h**, Transient PL of solution containing 10 mM of TBRb and 80 mM of TpAT-tFFO, compared to solution containing 10 mM of TBRb only.

Fig. 1b shows three-dimensional representations of the molecular structures of TpAT-tFFO and TpATtBu-tFFO. The triptycene linkers in these compounds enable a face-to-face alignment of the donor (dimethyl dihydrouridine) and acceptor groups (either diphenyl triazine or di-ter-butyl triazine), with the HOMO and LUMO localizing on the donor and acceptor groups, respectively. The "double-decker" structures are further expected to facilitate intermolecular π-π stacking at high concentrations. Consequently, these molecules are expected to allow intramolecular[32] as well as intermolecular CT transitions.

The concentration-dependent PL spectra of TpAT-tFFO solutions in a 2:1 by volume mixture of toluene and acetonitrile showed a shift in emission from blue fluorescence (482 nm) at <0.3 mM to green fluorescence (530 nm) at >30 mM (Figs. 1c and 1d). The gradual red-shift indicates the transition from monomer to excimer emission, with the saturation in spectral shift at 30 mM indicating that pure excimer state is reached at this concentration. According to our previous TD-DFT calculations for TpAT-tFFO[32], the energy level of the singlet CT state increases as the donor-acceptor separation increases. A reduction in energy by 0.23 eV corresponds to a decrease in separation from 4.72 Å (intramolecular CT state) to 4.00 Å (intermolecular CT state), driven by strong intermolecular electrostatic binding. Unlike the emission, the absorption spectra (spectrally resolved molar extinction coefficient) of TpAT-tFFO at concentrations of 0.1 mM and 10 mM are largely identical, with no red-shift of the CT absorption observed with increasing concentration (Fig. 1e). This indicates negligible intermolecular interactions of TpAT-tFFO in the ground state, even though strong intermolecular binding occurs upon excitation. Transient PL measurements indicate TADF is present across all concentrations tested here (Fig. 1f). The ratio of the photoluminescence quantum yield (PLQY) of the delayed and prompt emission ($\Phi_d/\Phi_p$) was higher for the excimer emission at high concentrations than for the monomer emission, while the ISC-RISC processes showed comparable rates (see Fig. 1g and Supplementary Table 1). Adding 10 mM 2,8-di-tert-butyl-5,11-bis(4-tert-butylphenyl)-6,12-diphenyltetracene (TBRb) to an 80 mM excimer solution of TpAT-tFFO yielded pure HF, with prompt and delayed fluorescence lifetimes of 25 ns and 172 ns, respectively (Fig. 1h) and no sign of residual excimer emission in this mixture (see Supplementary Fig. 1). The HF is facilitated by the very fast RISC on the excimer (rate constant of $8.2 \times 10^7$ $s^{-1}$) and also by rapid FRET.

Adding a supporting electrolyte to the hyperfluorescent solution, we fabricated two types of ECLDs in parallel-electrode (PE) configuration (Fig. 2a). The first, a glass-glass device, uses two ITO-coated glass substrates, separated by a 30-micrometer gap filled with the solution and had four square-shaped emissive surfaces measuring 4 $mm^2$ each (see Methods for detailed fabrication process). The intensity of light emitted through both substrates was equal due to the optically symmetric structure of the device; luminance values quoted in the following refer to the light emission to one side only. The second configuration, a glass-mirror device, includes a silver mirror with a passivation coating on the outer side of one of the glass substrates to reflect the emission back into and through the device and thus achieve unidirectional and therefore brighter emission through one of the substrates.

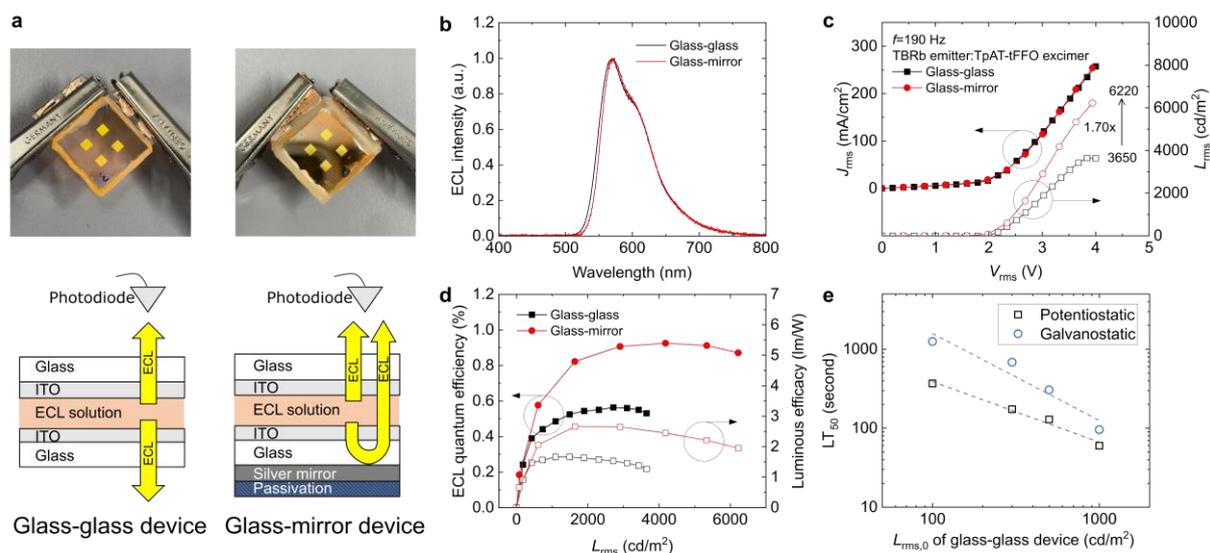

**Fig. 2. ECLDs operating based on the ECiHF mechanism using TpAT-tFFO as excimer-forming material and TBRb as terminal emitter. a**, Photographs and structures of glass-glass and glass-mirror parallel electrode (PE) device configurations with 4 pixels, each measuring 2 mm by 2 mm, operated simultaneously. **b**, ECL spectrum of each device. **c**, Root-mean-square (rms) current-voltage-luminescence (*JVL*) characteristics. **d**, Quantum efficiency and luminous efficacy. Data obtained under AC operation at a frequency of 190 Hz. **e**, Operational lifetime $LT_{50}$ of the glass-glass device under potentiostatic and galvanostatic operation, respectively. All data is based on emission to one side of the substrate only.

Fig. 2b shows the ECL spectra of these two devices; both spectra peak at 570 nm, clearly indicating that light emission is exclusively from TBRb. (By contrast, devices without TBRb showed direct ECL from the electrogenerated CT excimer; see Supplementary Fig. 2.) The glass-mirror devices showed a slightly reduced emission intensity at wavelengths shorter than the peak wavelength due to self-absorption by TBRb during light recycling. Fig. 2c shows the root-mean-square (rms) current density-voltage-luminance (*JVL*) characteristics of both devices under AC operation at a frequency of 190 Hz. Both devices showed very similar current-voltage characteristics due to their identical electrochemical nature. The glass-glass device achieved a luminance of 3,650 cd/m² at $V_{rms}$=4.0 V, while the glass-mirror device showed a 1.70-fold enhancement, reaching a maximum luminance of 6,220 cd/m². The ECL quantum efficiency ($\Phi_{ECL}$) and luminous efficacy (LE) shown reached maximum values of 0.56% and 1.66 lm/W, respectively, for the glass-glass device (only considering the emission in one direction), and 0.93% and 2.66 lm/W for the glass-mirror device (Fig. 2d). The glass-mirror device thus showed an improvement in luminance, LE, and $\Phi_{ECL}$, by 1.60 to 1.70-fold. This enhancement, while significant, was less than the theoretical improvement of 1.97 expected when considering the reflectivity of an ideal silver surface. The loss is likely due to differences in the ITO surface condition between the two devices as the UV-ozone treatment of ITO had to be reduced from the optimum duration of 15 minutes to 3 minutes post-coating of the silver mirror, which in turn will affect the faradaic process during device operation. Self-absorption by TBRb might further contribute to the lower-than-expected brightness in the glass-mirror device.

Fig. 2e shows estimates of $LT_{50}$ of our glass-glass ECLDs, obtained by measuring the time until the ECL intensity decreases to 50% of the maximum value when continuously operated under AC driving (see luminance vs time curves in Supplementary Fig. 3). Under potentiostatic (voltage-controlled) operation, the $LT_{50}$ values were 369 and 60 seconds at initial luminance levels of 100 cd/m$^2$ and 1,000 cd/m$^2$, respectively. Under galvanostatic (current-controlled) operation, the $LT_{50}$ values for these initial luminance levels improved to 1,252 and 96 seconds, respectively.

Table 1 summarizes the performance of the ECLDs in this study and compares them to prior works using the same PE configuration with ITO electrodes but different operating mechanisms, i.e. exciplex formation and ionic annihilation. Devices based on excimer formation outperformed annihilation-based devices in luminance as they use a more efficient bimolecular recombination process. The excimer-based device achieved a major further improvement in luminance, $\Phi_{ECL}$, and operational lifetime over the exciplex-based device. By further using reflection of back-emission, the glass-mirror device achieved a further increase in luminance, reaching an unprecedented ECL luminance of 6,220 cd/m$^2$. Galvanostatic operation enhanced the $LT_{50}$ lifetime relative to potentiostatic operation, due to the better charge carrier regulation which maintains ionic balance[18], and reduces local thermal stress due to local concentrations in charge flux at protruded regions of the slightly rough ITO surface[33]. As a result, the CT excimer-based device achieved an $LT_{50}$ value exceeding 1,200 seconds at an initial luminance of 100 cd/m$^2$, to our best knowledge, surpassing the current record of 1,000 seconds for a much dimmer ECLD using a Ru(bpy)$_3^{2+}$ luminophore with TiO$_2$ nanoparticles (where peak luminance was only 165 cd/m$^2$)[34]. We attribute these improvements mainly to two factors: (1) the presence of ECiHF with rapid ISC/RISC cycles on the CT excimer, and (2) the greater robustness of the excimer against chemical degradation at high-voltage and during prolonged operation.

**Table 1. Summary of characteristics of ECLDs with PE configuration, AC driving, and different operating mechanisms.** $M$ indicates a device with a mirror coating on one substrate side. $T$ indicates a device with a mesoporous TiO$_2$ electrode. Reference [29] reports $\Phi_{ECL}$ values of 0.35% and 0.52% for operation at 300 Hz and 100 Hz, respectively; this table shows the $\Phi_{ECL}$ value at 300 Hz, at which the excimer devices showed higher luminance. Reference [18] measured the lifetimes at an initial luminance of approximately 115 cd/m$^2$ and demonstrated a further improvement in lifetime for a floating bipolar electrode configuration. Reference [34] measured the lifetime at an initial luminance of 165 cd/m$^2$.

| Operating mechanism | Material | Medium | $L_{max}$ (cd/m$^2$) | $LE_{max}$ (lm/W) | $\Phi_{ECL}$ (%) | $LT_{50}$ (sec), $L_0$~100 cd/m$^2$ |
|---|---|---|---|---|---|---|
| CT excimer (this work) | TBRb & TpAT-tFFO | toluene-acetonitrile | 3650, 6220 ($M$) | 1.66, 2.66 ($M$) | 0.56, 0.93 ($M$) | 1252 |
| Exciplex[28,29] | TBRb, TAPC & TPBi | toluene-acetonitrile | 1250, 2260 ($T$) | 1.17, 2.06 ($T$) | 0.38 | 75 |
| Annihilation[28] | TBRb | toluene-acetonitrile | 390 | 0.74 | - | 26 |
| Annihilation[18,35] | Ru(bpy)$_3^{2+}$ | ionic liquid | 700 | 0.20 | - | 300 |
| Annihilation[34] | Ru(bpy)$_3^{2+}$ & TiO$_2$ nanoparticles | propylene carbonate | 165 | - | - | 1000 |

Next, we produced a simple ECLD display that leverages the enhanced brightness provided by ECiHF as well as the design freedom offered by ECL. Figure 3a shows the floating bipolar electrode (FBE)[18] configuration used for this display, with two parallel stripes of ITO on the top glass separated from 70-nm-thick gold electrodes on the bottom glass by a 30 µm tall gap that was filled with ECL solution. This architecture facilitates the display of calligraphy by fine-pattering of the gold; in the present example the gold electrodes were patterned with a resolution of 10 μm forming the words "Köln" (Cologne) and "京都" (Kyoto), the logos of the institutes involved in this study, and a tree branch motif. Applying AC voltage to the electrodes located along the edges of the substrate generated electrical double layers near the surfaces of the opposing electrodes, thereby triggering the ECL reaction. Figure 3b shows a photograph of the operating device, clearly displaying the calligraphy. Microscope images confirm that the light intensity profile closely follows the fine electrode pattern (Supplementary Fig. 4).

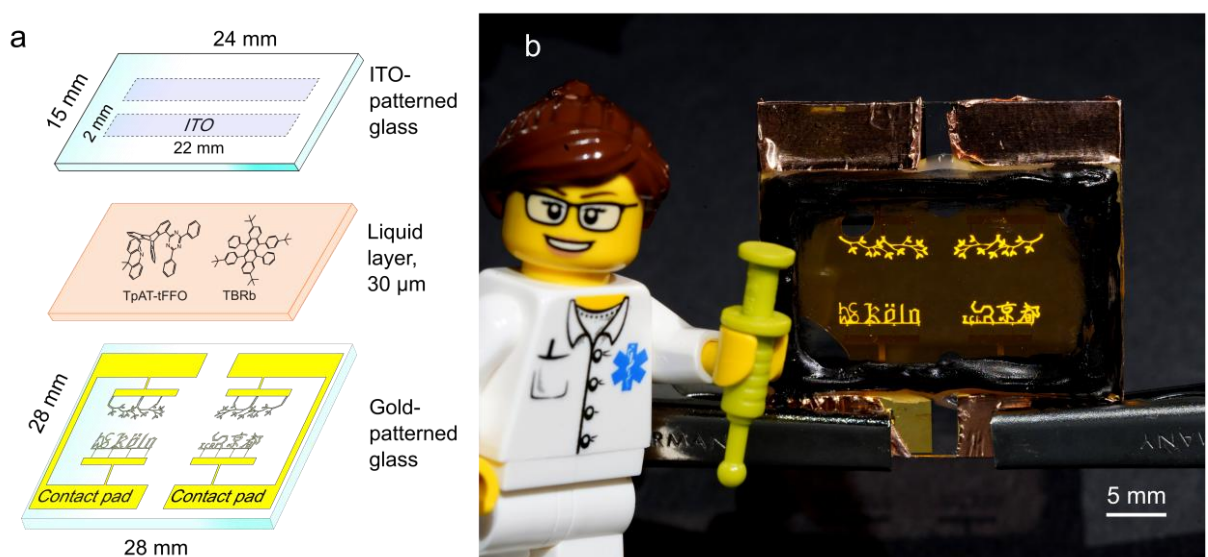

**Fig. 3. Calligraphic ECLD display. a**, Floating bipolar electrode (FBE) configuration featuring calligraphic gold electrodes with a minimum feature size of 10 μm and non-patterned rectangular ITO electrodes that are at a floating potential. The overlap of the gold and ITO electrodes defines a total emissive area of 11.4 mm². **b**, Photograph of device in operation with LEGO™ figure for size reference. Alligator clips supply AC voltage via contact pads on the left and right edges. The orange region in the center of the device is the area filled with the ECL solution.

We also studied ECiHF operation using the second excimer-forming material, TpATtBu-tFFO, which has a higher band gap (3.25 eV) than TpAT-tFFO (2.86 eV). Like TpAT-tFFO, TpATtBu-tFFO exhibited a concentration-dependent transition from monomer to excimer emission, shifting from deep-blue monomer emission at 442 nm to greenish-blue excimer emission at 484 nm (Fig. 4a). The excimer emission also exhibited strong delayed fluorescence, with higher

$\Phi_d/\Phi_p$ ratio than the monomer emission (see Supplementary Table 2 and Supplementary Fig. 5). The TpATtBu-tFFO excimer emission spectrum showed a stronger overlap with TBRb absorption than the TpAT-tFFO excimer (see Supplementary Fig. 6), thus potentially offering more efficient FRET and improved ECiHF performance.

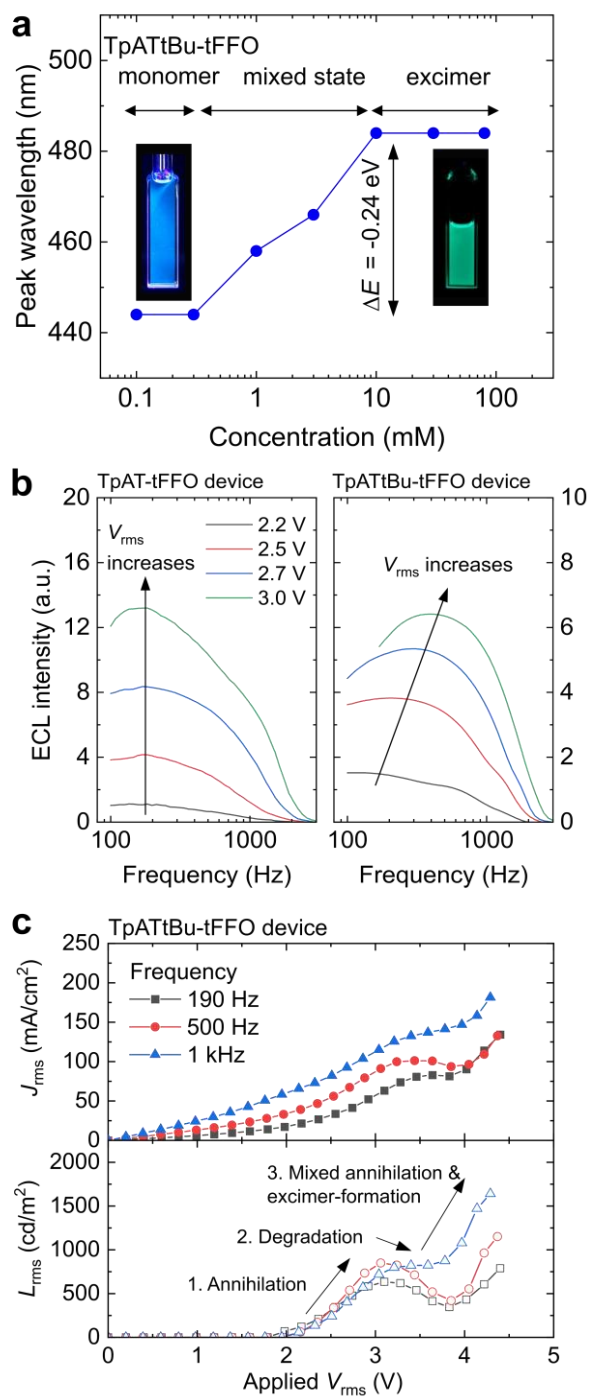

**Fig. 4. ECLDs based on excimer formation in TpATtBu-tFFO. a**, Shift in PL emission peak with concentration of TpATtBu-tFFO, demonstrating the transition from monomer to excimer emission. **b**, Frequency-dependent ECL intensity at different applied voltages, comparing

TpAT-tFFO and TpATtBu-tFFO devices. **c**, Rms *JVL* characteristics of the TpATtBu-tFFO device operating at 190 Hz, 500 Hz, and 1 kHz.

Figure 4b compares the frequency-dependent ECL intensity of TpAT-tFFO and TpATtBu-tFFO based devices in a glass-glass PE configuration at different applied voltages. While the TpAT-tFFO device showed no shift in optimal frequency ($f_{opt}$ = 190 Hz), the TpATtBu-tFFO device exhibited a gradual upward shift. Additionally, the TpATtBu-tFFO device displayed S-shaped *JVL* characteristics, i.e. a simultaneous decrease in current density and luminance at $V_{rms}$ > 3.0 V, followed by an increase in both at even higher voltages (Fig. 4c). A similar S-shaped *JV* curve was previously observed in an ECLD based on annihilation between TBRb ions[28] when rapid degradation of TBRb molecules occurred. However, in contrast to the TBRb devices, our TpATtBu-tFFO device showed a subsequent increase in luminance at higher voltage. These observations suggest that the TpATtBu-tFFO device operates via a mixed mechanism involving both ionic annihilation and excimer-formation. Below $V_{rms}$ = 3.0 V, annihilation dominates and causes rapid degradation of TBRb molecules. At higher voltage, the excimer-formation process becomes effective, which increases the luminance again as this process does not involve TBRb ions. Nevertheless, the two processes remain in competition, with their relative contributions depending on the operation frequency. Higher frequencies led to increased luminance, with a maximum of 1,640 cd/m² achieved at $f$ = 1 kHz.

To gain deeper insights into the differences between TpAT-tFFO and TpATtBu-tFFO devices, we performed a spectroelectrochemical analysis[36] as illustrated in Fig. 5a. This technique tracks changes in optical density (ΔOD) as light passes through the active area of the ECLD with and without an applied voltage. A positive ΔOD indicates the presence of electrogenerated ionic species, while a negative ΔOD reflects the depletion of neutral species due to faradaic processes. Because our analysis used a one-second acquisition time—much longer than a single AC cycle—it is particularly sensitive to the accumulation of ions involved in slower reaction pathways. In contrast, ions undergoing rapid processes are less detectable, as they rapidly revert to neutral states during acquisition.

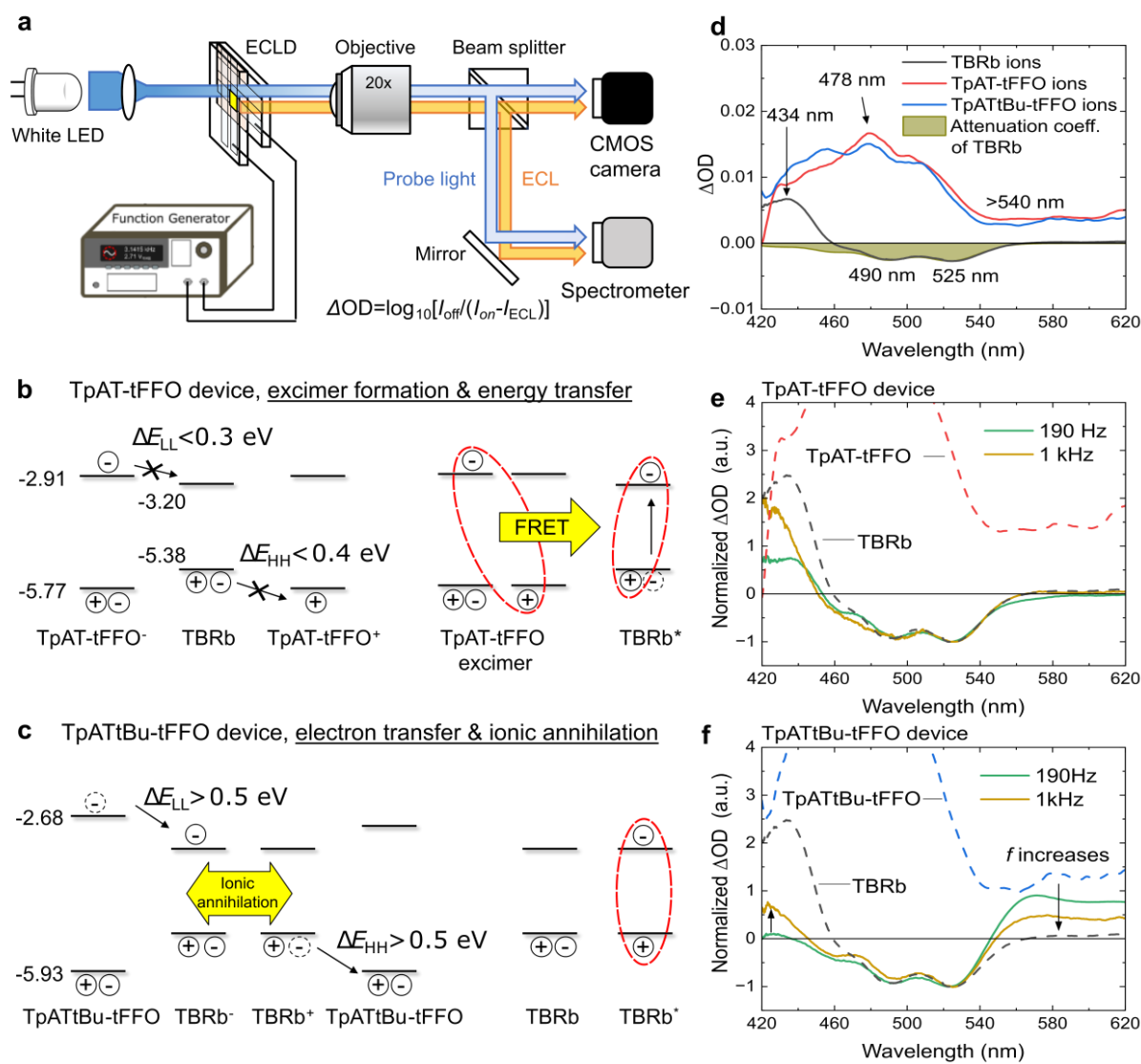

**Fig. 5. Absorption spectroelectrochemistry. a**, Schematic of the measurement setup. **b**, Schematic of the ECL process based on excimer-formation on TpAT-TFFO and subsequent energy-transfer to TBRb. $\Delta E_{HH}$ and $\Delta E_{LL}$ represent the offsets between the HOMO and LUMO levels, respectively. **c**, Schematic of electron-transfer ECL process between TpATtBu-tFFO and TBRb and subsequent ionic annihilation on TBRb. **d**, Measured absorption spectra of TBRb, TpAT-tFFO, TpATtBu-tFFO ions under DC operation. **e-f**, Spectroscopic analysis of optical density changes in TpAT-tFFO and TpATtBu-tFFO devices under AC operation at frequencies of 190 Hz and 1 kHz. Changes are relative to the off-state of each device (see Methods for details). The absence of TpATtBu-tFFO ion absorption at $\lambda = 478$ nm is likely because the ions responsible for this absorption rapidly disappear through faradaic processes.

Under an applied voltage, three reaction pathways can generate a TBRb exciton. First, a cation and an anion of the TpAT-tFFO or TpATtBu-tFFO monomers produced during the AC cycles can bind to form a CT excimer, which subsequently transfers energy to a TBRb molecule (Fig. 5b). Second, the ions of these monomers can transfer electrons to two separate TBRb molecules, creating a TBRb cation and anion, which subsequently annihilate to produce a TBRb exciton (Fig. 5c). The gaps in HOMO and LUMO levels ($\Delta E_{HH}$ and $\Delta E_{LL}$, respectively) between TpATtBu-tFFO and TBRb are both larger than those between TpAT-tFFO and TBRb, which may favor the second reaction pathway in the TpATtBu-tFFO system. Third, the TBRb molecule is ionized directly at the electrode surface although this route is less likely than the other two pathways due to the lower concentration of TBRb relative to TpAT-tFFO or TpATtBu-tFFO in devices.

Figure 5d shows the absorption spectra of TpAT-tFFO, TpATtBu-tFFO, and TBRb ions measured under DC operation (see Methods for details). Due to their structural similarity, TpAT-tFFO and TpATtBu-tFFO ions exhibited comparable absorption spectra, with a prominent peak at 478 nm and a broad absorption band extending beyond 540 nm. TBRb ions showed an absorption peak at 434 nm, while the negative $\Delta OD$ signals at 490 nm and 525 nm correspond to the loss of neutral TBRb molecules.

We then analyzed $\Delta OD$ under AC operation at $V_{rms} = 3.5$ V at operating frequencies of 190 Hz and 1 kHz. For the TpAT-tFFO device (Fig. 5e), TBRb ion absorption was detected near $\lambda = 434$ nm, but little TpAT-tFFO ion absorption was observed at either frequency as the ions rapidly formed excimers. In contrast, the TpATtBu-tFFO device (Fig. 5f) showed both TBRb ion absorption and a broad absorption band from TpATtBu-tFFO ions ($\lambda > 540$ nm). As the frequency increased, TpATtBu-tFFO ion absorption reduced in magnitude, while the absorption signature of TBRb ions increased. These findings confirm that the TpAT-tFFO device operates through a robust CT excimer mechanism, whereas the TpATtBu-tFFO device operates through a mixed mechanism involving both excimer formation and ionic annihilation. For the TpATtBu-tFFO device, the excimer formation process becomes more dominant at higher frequencies, consistent with the observed upward shift of optimal frequency with increasing voltage (Fig. 4b) and the enhanced luminance at higher-frequency operation (Fig. 4c).

Taken together, these observations indicate that since both $\Delta E_{HH}$ and $\Delta E_{LL}$ are less than 0.40 eV relative to TBRb for TpAT-tFFO, fast excimer formation outcompetes electron transfer processes in the mixture. By contrast, for TpATtBu-tFFO, $\Delta E_{HH}$ and $\Delta E_{LL}$ both exceed 0.50 eV, which promotes electron transfer processes that generate TBRb ions, particularly under a low-frequency AC operation.

In summary, we explored the mechanism of ECiHF and demonstrated significant improvements in luminance, efficiency, and brightness of ECLDs, enhancing their potential for lighting and display applications. The TADF materials TpAT-tFFO and TpATtBu-tFFO formed CT excimers exhibiting strong delayed fluorescence at concentrations above 30 mM. Rapid RISC on the TpAT-tFFO excimer, followed by FRET to fluorescent TBRb, resulted in efficient hyperfluorescence. An ECLD fabricated with two ITO-coated glass substrates achieved a maximum luminance of 3,650 cd/m² in each direction, which increased to 6,220 cd/m² upon adding a silver mirror to retro-reflect emission from one side of the device. Additionally, for an initial luminance of 100 cd/m², the $LT_{50}$ values exceeded 6 minutes under potentiostatic operation and 20 minutes under galvanostatic operation—representing the most stable ECLD performance reported to date. Combined with an FBE device structure, the enhanced brightness enabled by ECiHF, facilitated the realization of a calligraphic display with 10 μm resolution. These advances in performance are poised to open up opportunities for new device architectures and, in turn, new applications of this unique class of fluid-state light-emitting devices.

To shed light on the physics of the ECiHF process and to establish guidelines for material selection and performance optimization, we investigated the impact of energy level alignment between the excimer-forming materials and terminal emitter. We found that $\Delta E_{HH}$ and $\Delta E_{LL}$ below 0.4 eV is beneficial for efficient ECiHF. The device using TpAT-tFFO, where both gaps are below this threshold, exhibited a monotonic increase in *JVL*, an optimal operating frequency (190 Hz) independent of applied voltage, and a robust excimer-formation process unaffected by operating frequency. In contrast, the device using TpATtBu-tFFO, where $\Delta E_{HH}$ and $\Delta E_{LL}$ exceed 0.5 eV, showed S-shaped and frequency dependent *JVL* curves and an upward shift in the optimal frequency with increasing voltage, both likely arising from a mixed operating mechanism involving excimer formation and ionic annihilation. Spectroelectrochemical analysis further revealed that at higher frequencies the predominant operating mechanism in the TpATtBu-tFFO device is excimer formation and subsequent energy transfer, whereas electron transfers and subsequent ionic annihilation appear to dominate at lower frequencies.

## Methods

### Calculations

Molecular structures were optimized using density functional theory (DFT) and HOMO and LUMO distributions were calculated by time-dependent DFT (TD-DFT). Both calculations were performed at the PBE0/6-31G(d) level of theory using the polarizable continuum model in toluene within the Gaussian 16 software package[37]. A dielectric constant of 18.40 was used for the mixture of toluene and acetonitrile.

### Material characterization

$^1$H and $^{13}$C NMR spectra were obtained with JEOL ECS400. CDCl$_3$ was used as a deuterated solvent and all measurements were performed at ambient temperature. Chemical shifts were reported in δ (ppm), using tetramethylsilane as internal standards. Atmospheric pressure chemical ionization (APCI) mass spectra were measured with a timsTOF (Bruker).

### Preparation of solutions

For PL samples, 80 mM TpAT-tFFO and 80 mM TpATtBu-tFFO were each weighed into separate vials. A mixture of anhydrous toluene and anhydrous acetonitrile in 2:1 by volume was pipetted into each vial. The solutions were heated at 60°C on a hot plate for 30 minutes. These 80 mM solutions were subsequently diluted to prepare concentrations down to 0.1 mM. For the hyperfluorescence PL sample, 10 mM TBRb was weighed into a vial, and then the 80 mM TpAT-tFFO solution was pipetted in. For ECL solutions, we first prepared a solution containing 10 mM TBRb and 100 mM supporting electrolyte (tetrabutylammonium hexafluorophosphate). Next, 80 mM TpAT-tFFO and 80 mM TpATtBu-tFFO were each weighed into separate vials. The TBRb solution was subsequently pipetted to each vial, followed by 30 minutes of heating at 60°C.

### PL and absorption measurement

The solutions were argon bubbled for longer than 30 minutes to remove residual oxygen. 700 μL quartz cuvettes were used for the PL characterization. The PL spectra and transient PL of the solutions were measured using a fluorescence lifetime spectrometer (FluoTime 250, PicoQuant) paired with a picosecond diode laser (λ=373 nm). UV-Vis absorption for 0.1 mM and 10 mM was measured with a UV-Vis spectrometer (Cary 50, Varian). The HOMO energy levels of TpAT-tFFO and TpATtBu-tFFO were determined utilizing photoelectron yield spectroscopy (AC-3, Riken Keiki). The LUMO energy levels were calculated from the HOMO energy levels and the optical band gaps. The HOMO and LUMO levels of TBRb are taken from Reference [38]. The PLQY values were determined by using an absolute PLQY spectrometer (C9920-02, Hamamatsu Photonics).

### Photolithography and lift-off processes

A glass substrate, measuring 96 mm by 96 mm, was cleaned with acetone and isopropanol. After drying, the substrate underwent oxygen plasma treatment for 3 minutes. A layer of positive photoresist (AR-P 3740, Allresist GmbH) was spin-coated onto the substrate at 4000 rpm for 60 seconds, followed by a soft bake at 90 °C for 1 minute. The prepared substrate was aligned and loaded into a laser writer (Picomaster 150, Raith GmbH) for pattern transfer via selective light exposure. Following exposure, the photoresist was developed (AR 300-26, Allresist GmbH, diluted 1:3 with purified water) and dried and. the substrate was placed in a vacuum chamber. A thin adhesion promoting layer of chromium and a 70-nm-thick gold layer were deposited at a base pressure of approximately 10$^{-7}$ mbar. Finally, the substrate was immersed in an acetone in an ultrasonic bath for 30 minutes. This lift-off step removed the photoresist and any gold deposited onto the resist, leaving only the desired gold pattern on the substrate.

**Device fabrication**

The fabrication of each ECLD with PE structure used a pair of rectangular glass substrates measuring 24 mm by 15 mm coated with two stripes of 2 mm-wide ITO. The substrates were cleaned using a detergent solution (2% Hellmanex III in Milli-Q water) followed by isopropyl alcohol. After drying, the ITO surface was treated with UV-ozone for 15 minutes. The substrate pair was bonded with NOA 68 resin (Norland Products), which was mixed with 30 μm-sized microbeads (Sigma-Aldrich), to have four 4 mm$^2$-size cross sections of ITO, and cured with UV light while being pressed using a custom-made holder (Detailed information can be found in ref. [28]). The edges of the bonded substrates remained open, maintaining a 30 μm gap, as the resin was applied only as droplets to the edges and corners. The bonded substrates were then transferred to a nitrogen-filled glove box, where 6.5 μL of the ECL solution was pipetted into the gap between the ITO substrates to fill the structure. The open edges were sealed with 3035BT resin (Threebond International) and cured with UV light, resulting in a production of glass-glass devices. For the glass-mirror device, after cleaning the substrate, a 120 nm-thick silver layer followed by a 50 nm-thick aluminum oxide layer was deposited onto the bare-glass side of the ITO substrates using vacuum evaporation and atomic layer deposition techniques, respectively. After coating, the substrates underwent the cleaning process again. The ITO surface was treated with UV-ozone for 3 minutes. A pair of silver-coated and uncoated ITO substrates was then used to fabricate each glass-mirror device following the steps described above.

For the fabrication of the ECLD in FBE configuration, the gold-patterned glass substrate was diced into nine pieces, each measuring 28 mm by 28 mm, using a diamond cutter. A diced substrate was precisely aligned with a rectangular glass substrate containing two stripes of 2 mm-wide ITO using a custom-made holder. The glasses were bonded together with NOA 68 resin containing microbeads, leaving the edges of the bonded substrates open and setting the separation between the substrates to 30 μm. The bonded substrates were transferred to the glove box, where 10 μL of the ECL solution was pipetted into the gap between the substrates. Finally, the open edges were sealed with 3035BT resin.

**Device characterization**

The device was mounted on a sample holder positioned at the center of a 55 cm by 55 cm sized dark box. *JVL* was characterized by a step-wise increase in the amplitude of sinusoidal voltage signal generated by a function generator (33220A, Agilent Technologies). The actual $V_{rms}$ and $I_{rms}$ values were recorded by a power analyzer (GPM-8213, GW Instek) during the voltage sweep. A silicon photodiode (PDA100A2, Thorlabs) positioned 168 mm from the device measured the photocurrent in AC mode. Spectral data were collected using a fiber-coupled spectrometer (Ocean HDX, Ocean Insight). A Lambertian distribution was assumed in the calculation of $\Phi_{ECL}$. Operational lifetime was analyzed using the same setup, with the photocurrent recorded every second during the extended operation. The voltage amplitude was controlled in potentiostatic mode, while the amplitude was finely adjusted to ensure the rms current remained within an error range of 1.0 μA in galvanostatic mode. Custom software automized all measurements.

## Absorption spectroelectrochemistry

Spectroelectrochemical analysis was performed using a custom-built inverted microscope setup (Nikon Eclipse Ti2) equipped with 20x extra-long working distance air objective. The sample was illuminated by a white LED source (pE-4000, CoolLED). A beam splitter and a digital camera (C13440 Orca-Flash 4.0, Hamamatsu) ensured precise alignment of the observation area within the active region of the device. The transmitted light spectrum was recorded using a spectrometer (Andor Shamrock 500i) with a 1-second acquisition time. First, the transmission of the white LED was measured with the device turned off ($I_{off}$ in Fig. 5a), followed by a measurement with the device turned on ($I_{on}$). Lastly, the ECL intensity was measured with the white LED switched off ($I_{ECL}$). The absorption of TpAT-tFFO and TpATtBu-tFFO ions was measured by applying a 3.5 V DC voltage to each TpAT-tFFO and TpATtBu-tFFO device used in Fig. 5e-f. The absorption of TBRb ions was measured using a device with a solution containing 10 mM TBRb with a supporting electrolyte, with application of 3.0 V DC voltage.

## Acknowledgements
The authors thank Matthias König for performing macro-photography of the calligraphic ECLD display. This work was financially supported by the Alexander von Humboldt Foundation (Humboldt-Professorship to M.C.G.), JSPS KAKENHI (JP20H05840, Grant-in-Aid for Transformative Research Areas, "Dynamic Exciton"; and JP23KJ1253), JST CREST (Grant No: JPMJCR2431), the JSPS Core-to-Core Program (JPJSCCA20220004), the International Collaborative Research Program of Institute for Chemical Research, Kyoto University (2024-126) and the Deutsche Forschungsgemeinschaft (Research Training Group "TIDE", RTG2591). A.P. acknowledges funding from the European Molecular Biology Organization through the EMBO Postdoctoral Fellowship (675-2022). Quantum chemical calculations were performed on the SuperComputer System, Institute for Chemical Research, Kyoto University.

## Author contributions
C.-K.M., H.K., and M.C.G. conceived the project. C.-K.M. conducted PL measurements, manufactured ECLDs, and characterized them. Y.Y. and Y. K. conducted PL measurements and computations for the tFFO-based molecules. S.F. synthesized TADF emitters. A.P. contributed to the spectroelectrochemical analysis. J.F.B. built the device characterization setup. N.P. contributed to photolithography. C.-K.M., H.K., and M.C.G. mainly wrote the manuscript.

## Competing interests
The authors declare no competing interests.

## Data availability
All primary data for all figures and extended data figures are available from the corresponding author upon request.

## References

1   Ma, C., Cao, Y., Gou, X. & Zhu, J.-J. Recent progress in electrochemiluminescence sensing and imaging. *Anal. Chem.* **92**, 431-454 (2019).

2	Muzyka, K. Current trends in the development of the electrochemiluminescent immunosensors. *Biosens. Bioelectron.* **54**, 393-407 (2014).

3	Miao, W. Electrogenerated chemiluminescence and its biorelated applications. *Chem. Rev.* **108**, 2506-2553 (2008).

4	Hao, N. & Wang, K. Recent development of electrochemiluminescence sensors for food analysis. *Anal. Bioanal. Chem.* **408**, 7035-7048 (2016).

5	Shen, Y., Gao, X., Lu, H.-J., Nie, C. & Wang, J. Electrochemiluminescence-based innovative sensors for monitoring the residual levels of heavy metal ions in environment-related matrices. *Coord. Chem. Rev.* **476**, 214927 (2023).

6	Kuik, M. *et al.* 25th anniversary article: charge transport and recombination in polymer light-emitting diodes. *Adv. Mater.* **26**, 512-531 (2014).

7	Meier, S. B. *et al.* Light-emitting electrochemical cells: recent progress and future prospects. *Mater. Today* **17**, 217-223 (2014).

8	Tang, S. & Edman, L. Light-Emitting Electrochemical Cells: A Review on Recent Progress. *Top. Curr. Chem.* **374** (2016).

9	Okumura, R., Takamatsu, S., Iwase, E., Matsumoto, K. & Shimoyama, I. in Proceedings of the 2009 *IEEE 22$^{nd}$ International Conference on Micro Electro Mechanical Systems (MEMS)*, Sorrento, Italy, 2009, 947-950.

10	Kasahara, T. *et al.* Recent advances in research and development of microfluidic organic light-emitting devices. *JPST* **30**, 467-474 (2017).

11	Moon, H. C., Lodge, T. P. & Frisbie, C. D. Solution-processable electrochemiluminescent ion gels for flexible, low-voltage, emissive displays on plastic. *J. Am. Chem. Soc.* **136**, 3705-3712 (2014).

12	Kim, S. *et al.* All-Printed Electrically Driven Lighting via Electrochemiluminescence. *Adv. Mater. Tech.* 2302190 (2024).

13	Cho, K. G. *et al.* Light-Emitting Devices Based on Electrochemiluminescence Gels. *Adv. Funct. Mater.* **30** (2020).

14	Nobeshima, T., Morimoto, T., Nakamura, K. & Kobayashi, N. Advantage of an AC-driven electrochemiluminescent cell containing a $Ru(bpy)_3^{2+}$ complex for quick response and high efficiency. *J. Mater. Chem.* **20** (2010).

15	Cao, Z., Shu, Y., Qin, H., Su, B. & Peng, X. Quantum dots with highly efficient, stable, and multicolor electrochemiluminescence. *ACS Cent. Sci.* **6**, 1129-1137 (2020).

16	Chu, K., Ding, Z. & Zysman-Colman, E. Materials for Electrochemiluminescence: TADF, Hydrogen-Bonding, and Aggregation- and Crystallization-Induced Emission Luminophores. *Chemistry* **29**, e202301504 (2023).

17	Kwon, D. K. & Myoung, J. M. Wearable and Semitransparent Pressure-Sensitive Light-Emitting Sensor Based on Electrochemiluminescence. *ACS Nano* **14**, 8716-8723 (2020).

18	Yee, H. *et al.* Extending the Operational Lifetime of Electrochemiluminescence Devices by Installing a Floating Bipolar Electrode. *Small* **20**, 2307190 (2024).

19	Laser, D. & Bard, A. J. Electrogenerated Chemiluminescence: XXIII. On the Operation and Lifetime of ECL Devices. *J. Electrochem. Soc.* **122**, 632 (1975).

20	Uoyama, H., Goushi, K., Shizu, K., Nomura, H. & Adachi, C. Highly efficient organic light-emitting diodes from delayed fluorescence. *Nature* **492**, 234-238 (2012).

21	Kaji, H. *et al.* Purely organic electroluminescent material realizing 100% conversion from electricity to light. *Nat. Comm.* **6**, 8476 (2015).

22	Ishimatsu, R. *et al.* Electrogenerated chemiluminescence of donor–acceptor molecules with thermally activated delayed fluorescence. *Angew. Chem. Int. Ed.* **53**, 6993-6996 (2014).

23	Huang, P. *et al.* Polymer electrochemiluminescence featuring thermally activated delayed fluorescence. *Chem. Phys. Chem.* **22**, 726-732 (2021).

24  Huang, P. *et al.* Studies on annihilation and coreactant electrochemiluminescence of thermally activated delayed fluorescent molecules in organic medium. *Molecules* **27**, 7457 (2022).
25  Ishimatsu, R. *et al.* Solvent effect on thermally activated delayed fluorescence by 1, 2, 3, 5-tetrakis (carbazol-9-yl)-4, 6-dicyanobenzene. *J. Phys. Chem. A* **117**, 5607-5612 (2013).
26  Legaspi, C. M. *et al.* Rigidity and polarity effects on the electronic properties of two deep blue delayed fluorescence emitters. *J. Phys. Chem. C* **122**, 11961-11972 (2018).
27  Nakanotani, H. *et al.* High-efficiency organic light-emitting diodes with fluorescent emitters. *Nat. Comm.* **5**, 4016 (2014).
28  Moon, C. K., Butscher, J. F. & Gather, M. C. An Exciplex-Based Light-Emission Pathway for Solution-State Electrochemiluminescent Devices. *Adv. Mater.* **35**, 2302544 (2023).
29  Moon, C. K. & Gather, M. C. Absolute quantum efficiency measurements of electrochemiluminescent devices through electrical impedance spectroscopy. *Adv. Optic. Mater.* **12**, 2401253 (2024).
30  Zhao, B. *et al.* Highly efficient red OLEDs using DCJTB as the dopant and delayed fluorescent exciplex as the host. *Sci. Rep.* **5**, 10697 (2015).
31  Kim, K.-H., Yoo, S.-J. & Kim, J.-J. Boosting triplet harvest by reducing nonradiative transition of exciplex toward fluorescent organic light-emitting diodes with 100% internal quantum efficiency. *Chem. Mater.* **28**, 1936-1941 (2016).
32  Wada, Y., Nakagawa, H., Matsumoto, S., Wakisaka, Y. & Kaji, H. Organic light emitters exhibiting very fast reverse intersystem crossing. *Nat. Photon.* **14**, 643-649 (2020).
33  Ko, E.-S. *et al.* Pulsed Driving Methods for Enhancing the Stability of Electrochemiluminescence Devices. *ACS Photon.* **5**, 3723-3730 (2018).
34  Tsuneyasu, S., Ichihara, K., Nakamura, K. & Kobayashi, N. Why were alternating-current-driven electrochemiluminescence properties from Ru(bpy)$_3^{2+}$ dramatically improved by the addition of titanium dioxide nanoparticles? *Phys. Chem. Chem. Phys.* **18**, 16317-16324 (2016).
35  Kong, S. H., Lee, J. I., Kim, S. & Kang, M. S. Light-Emitting Devices Based on Electrochemiluminescence: Comparison to Traditional Light-Emitting Electrochemical Cells. *ACS Photon.* **5**, 267-277 (2017).
36  Yasuji, K., Sakanoue, T., Yonekawa, F. & Kanemoto, K. Visualizing electroluminescence process in light-emitting electrochemical cells. *Nat. Comm.* **14**, 992 (2023).
37  Frisch, M. J. *et al.* Gaussian 16, revision C.01 (Gaussian Inc., 2016).
38  Liu, T.-H. *et al.* Highly efficient yellow and white organic electroluminescent devices doped with 2, 8-di (t-butyl)-5, 11-di [4-(t-butyl) phenyl]-6, 12-diphenylnaphthacene. *Appl. Phys. Lett.* **85**, 4304-4306 (2004).

Supplementary Information

# Electrochemically induced hyperfluorescence based on the formation of charge-transfer excimers

Chang-Ki Moon[1,2], Yuka Yasuda[3], Yu Kusakabe[3], Anna Popczyk[1], Shohei Fukushima[3], Julian Butscher[1], Nachiket Pathak[1], Hironori Kaji[3]*, Malte C. Gather[1,2]*

[1]Humboldt Centre for Nano- and Biophotonics, Institute for Light and Matter, Department of Chemistry and Biochemistry, University of Cologne, Greinstr. 4-6, 50939 Köln, Germany

[2]Organic Semiconductor Centre, School of Physics and Astronomy, University of St Andrews, North Haugh, St Andrews KY16 9SS, United Kingdom

[3]Institute for Chemical Research, Kyoto University, Uji, Kyoto, Japan

**Supplementary Table 1. TADF parameters of TpAT-tFFO at various concentrations in mixed toluene and acetonitrile solutions.** The lifetime parameters are extracted from transient PL curves using the equation, $I(t) = I_p \exp(-t/\tau_p) + I_d \exp(-t/\tau_d)$. The analysis of $k_{ISC}$ and $k_{RISC}$ used the method described in Methods section of reference[1].

| Concentration (mM) | State | $I_p$ | $\tau_p$ (ns) | $I_d$ | $\tau_d$ (ns) | $\Phi_d/\Phi_p$ | $\Phi_{PL}$ | $k_{ISC}$ ($10^7$/s) | $k_{RISC}$ ($10^7$/s) | $k_{ISC}k_{RISC}$ ($10^{14}$/s) |
|---|---|---|---|---|---|---|---|---|---|---|
| 0.1 | Monomer | 0.837 | 17.3 | 0.144 | 977 | 9.73 | 0.84 | 4.02 | 1.42 | 5.70 |
| 0.3 | Monomer | 0.883 | 18.3 | 0.128 | 1150 | 9.16 | 0.84 | 4.03 | 1.07 | 4.31 |
| 1 | Mixed | 0.837 | 20.2 | 0.158 | 928 | 8.65 | - | - | - | - |
| 3 | Mixed | 0.831 | 19.6 | 0.149 | 492 | 4.52 | - | - | - | - |
| 10 | Mixed | 0.838 | 20.4 | 0.158 | 355 | 3.27 | - | - | - | - |
| 30 | Excimer | 0.837 | 20.7 | 0.157 | 395 | 3.58 | 0.27 | 2.70 | 1.62 | 4.37 |
| 80 | Excimer | 0.794 | 18.6 | 0.131 | 1770 | 15.7 | 0.27 | 3.76 | 1.20 | 4.51 |

**Supplementary Table 2. TADF parameters of TpATtBu-tFFO at various concentrations in mixed toluene and acetonitrile solutions.** The lifetime parameters are extracted from transient PL curves using the equation, $I(t) = I_p \exp(-t/\tau_p) + I_d \exp(-t/\tau_d)$. We found that TpATtBu-tFFO is a TADF molecule with particularly high sensitivity to oxygen. Prior to the PL measurement, each solution was purged with argon for 30 minutes to minimize oxygen interference. However, trace oxygen may still be present in the solution and quench triplet emission from TpATtBu-tFFO during the measurement, potential leading to errors in the ISC and RISC parameters of TpATtBu-tFFO reported here.

| Concentration (mM) | State | $I_p$ | $\tau_p$ (ns) | $I_d$ | $\tau_d$ (ns) | $\Phi_d/\Phi_p$ | $\Phi_{PL}$ | $k_{ISC}$ ($10^7$/s) | $k_{RISC}$ ($10^7$/s) | $k_{ISC}k_{RISC}$ ($10^{14}$/s) |
|---|---|---|---|---|---|---|---|---|---|---|
| 0.1 | Monomer | 0.744 | 2.61 | 0.160 | 44.8 | 3.68 | 0.52 | 18.5 | 16.9 | 313 |
| 0.3 | Monomer | 0.737 | 2.95 | 0.144 | 53.5 | 3.90 | 0.52 | 16.7 | 14.8 | 247 |
| 1 | Mixed | 0.667 | 3.18 | 0.193 | 56.4 | 5.14 | - | - | - | - |
| 3 | Mixed | 0.725 | 3.44 | 0.254 | 64.2 | 6.53 | - | - | - | - |
| 10 | Excimer | 0.895 | 2.44 | 0.082 | 327 | 12.3 | 0.50 | 34.1 | 4.53 | 154 |
| 30 | Excimer | 0.835 | 3.06 | 0.138 | 328 | 17.6 | 0.50 | 24.3 | 7.21 | 175 |
| 80 | Excimer | 0.855 | 3.12 | 0.158 | 379 | 22.4 | 0.50 | 23.1 | 8.23 | 190 |

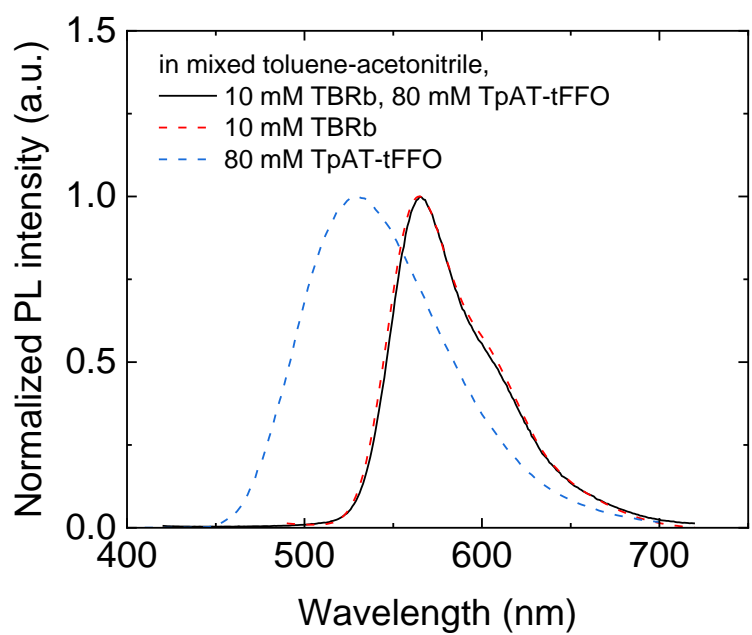

**Supplementary Fig. 1. PL spectra of the mixed toluene-acetonitrile solution containing 10 mM of TBRb and 80 mM of TpAT-tFFO, both together and individually**. Both TBRb and TpAT-tFFO molecules are excited by $\lambda = 373$ nm light. In the mixture of TBRb and TpAT-tFFO, the emission is solely from TBRb due to rapid energy transfer from TpAT-tFFO excimers to TBRb molecules.

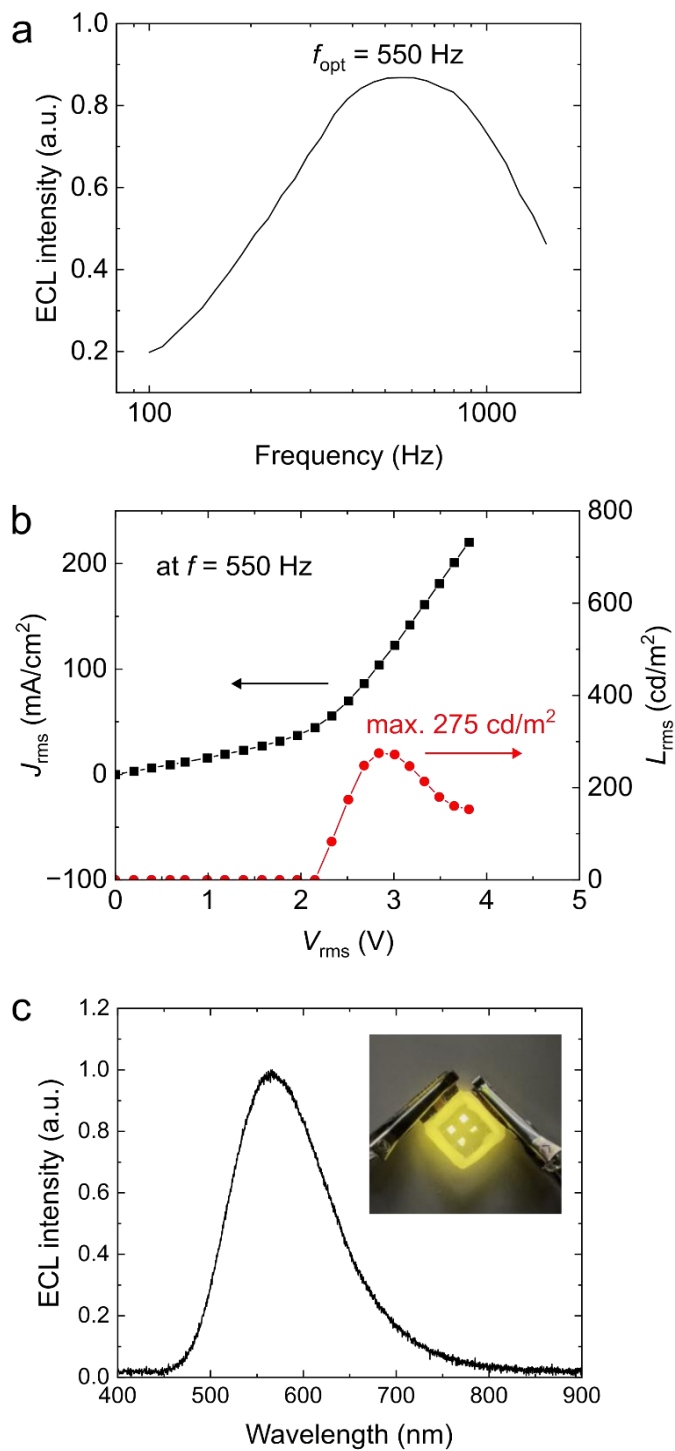

**Supplementary Fig. 2. CT excimer emission ECLD**. Parallel electrode (PE) structure device with two ITO-coated glass substrates and a mixed acetone and acetonitrile ECL solution containing 80 mM TpAT-tFFO and 100 mM electrolyte. **a**, Frequency-dependent ECL intensity measured under AC operation. **b**, Current-voltage-luminance characteristics measured at a frequency of 500 Hz. **c**, ECL spectrum and photograph of the device in operation.

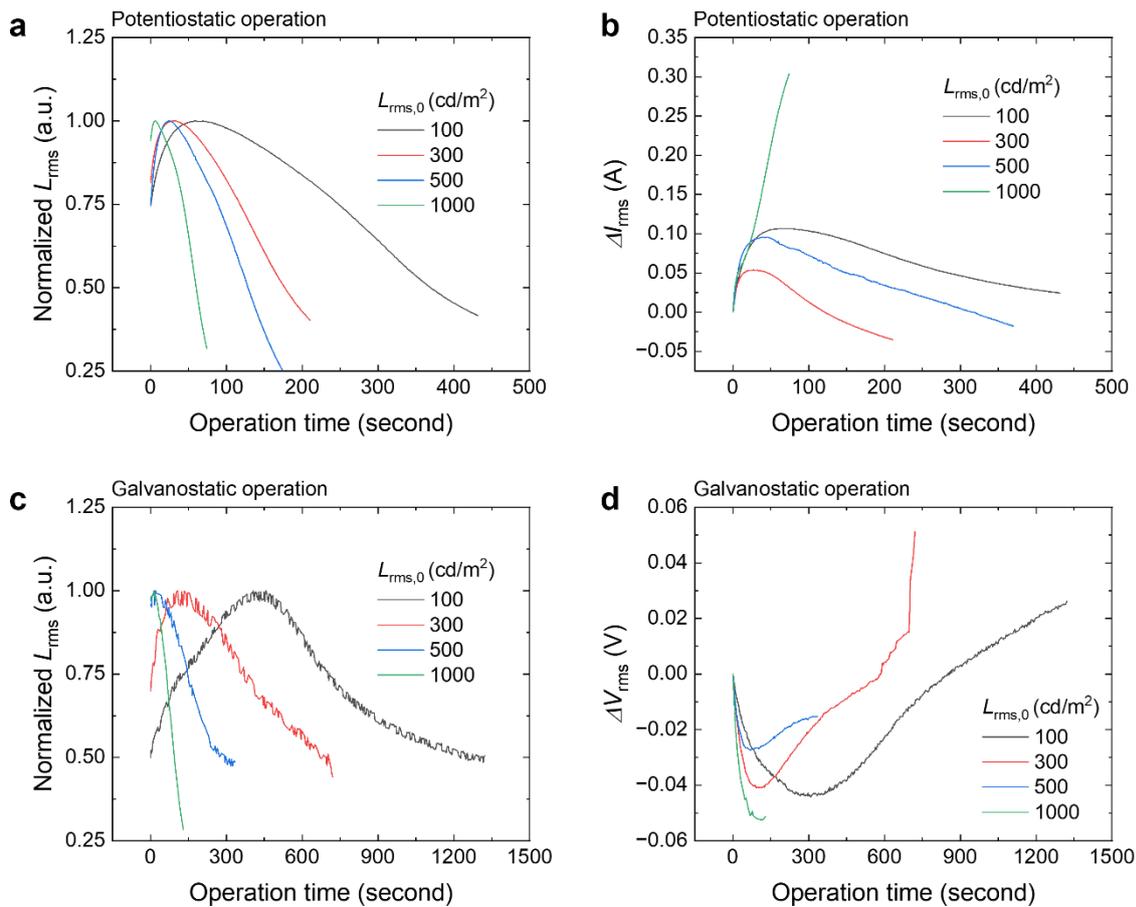

**Supplementary Fig. 3. Measurement of operational lifetime of the TpAT-tFFO device. a**, Luminance over time under potentiostatic operation at various initial luminance values and **b**, corresponding changes in the rms current. **c**, Luminance over time under galvanostatic operations and **d**, changes in the voltage applied to maintain the preset current. Voltage application leads to re-distribution of molecules and ions within the liquid layer. This in turn results in an initial increase in luminance along with an increase in device current under potentiostatic operation and decrease in voltage under the galvanostatic operation. The $LT_{50}$ value was estimated as the time when the brightness decreased by half of the maximum value for each operation condition.

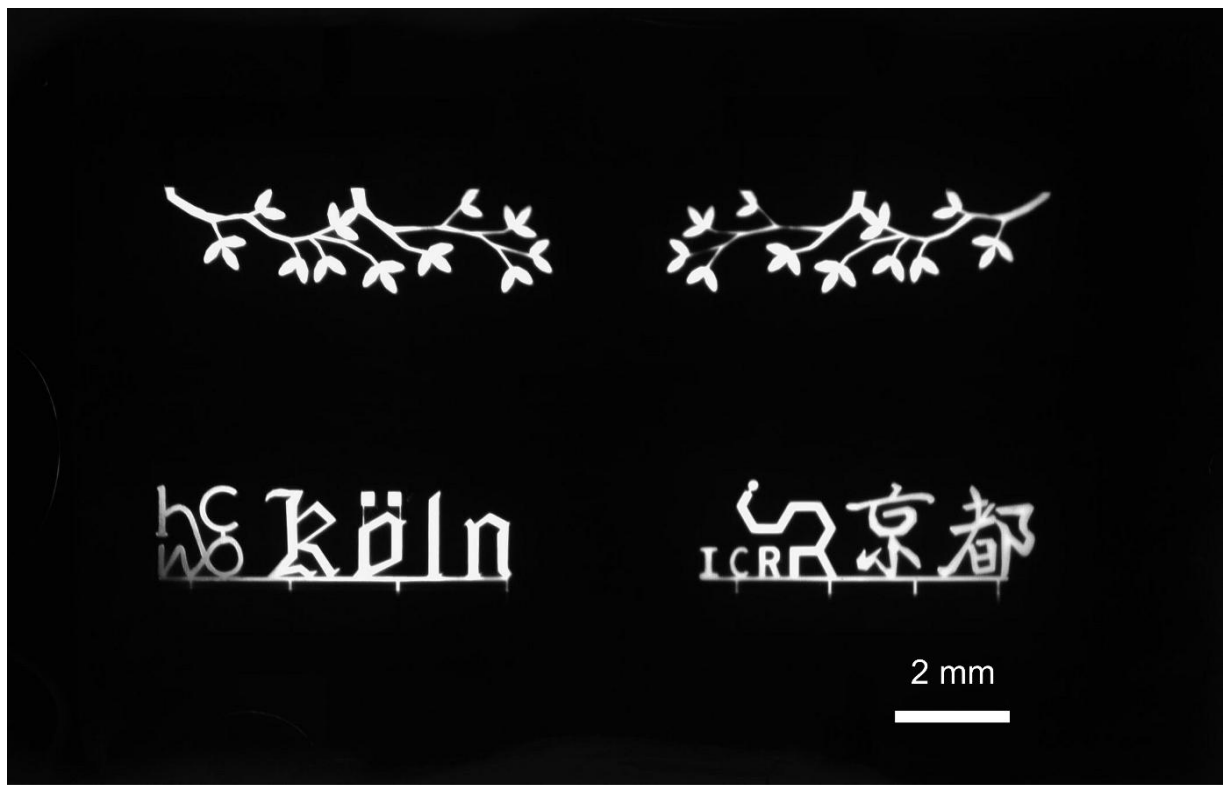

**Supplementary Fig. 4.** High-resolution microscope image showing the light intensity profile in greyscale from a calligraphic ECLD display configured in floating bipolar electrode (FBE) configuration.

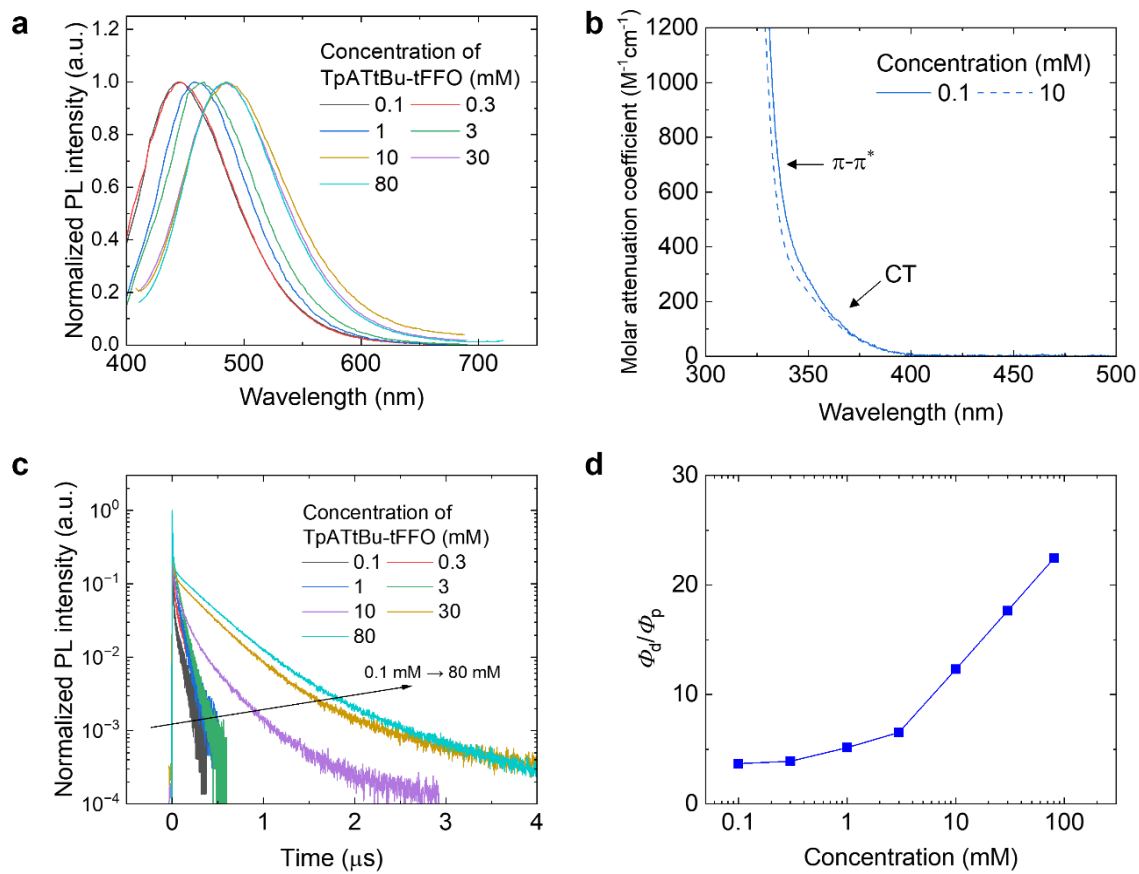

**Supplementary Fig. 5. Photoluminescence characteristics of TpATtBu-tFFO solutions. a**, PL spectra, **b**, molar attenuation coefficient, **c**, transient PL, and **d**, PLQY ratio of delayed fluorescence to prompt fluorescence of TpATtBu-tFFO for various concentrations in mixed toluene-acetonitrile solutions (2:1 by volume).

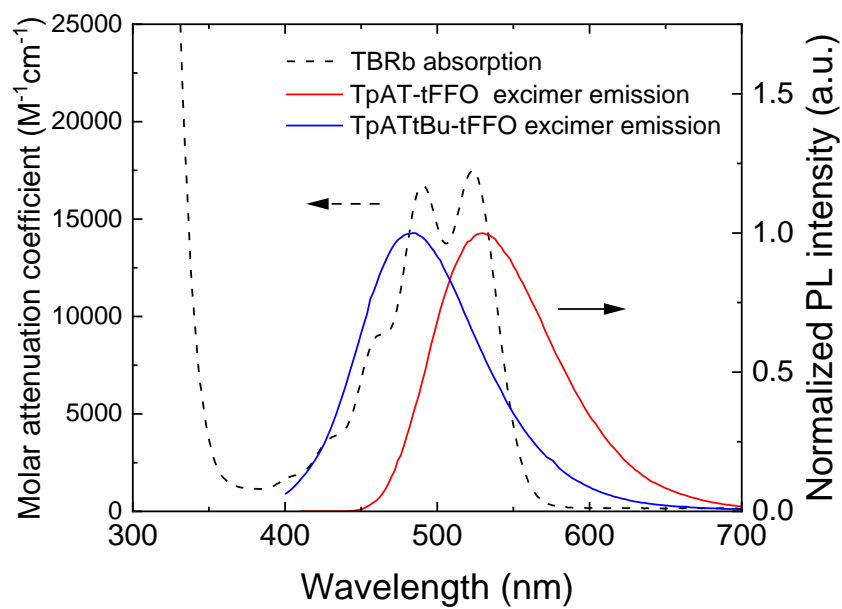

**Supplementary Fig. 6. Absorption spectrum of TBRb and emission spectra of the TpAT-tFFO and TpATtBu-tFFO excimers.**

## Synthesis of TpAT-tFFO

TpAT-tFFO was prepared according to reference[1].

## Synthesis of TpATtBu-tFFO

The synthesis scheme of TpATtBu-tFFO is summarized in Supplementary Scheme 1 and 2. The starting material S1 was prepared according to reference[1].

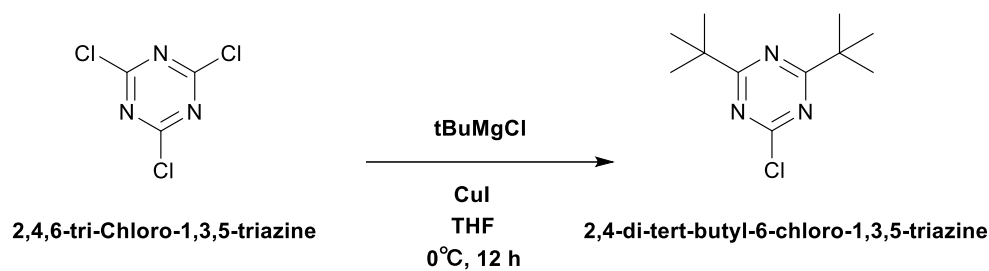

**Supplementary Scheme 1.** Synthesis of 2,4-di-tert-butyl-6-chloro-1,3,5-triazine.

In a 100-mL two neck round bottom flask, 2,4,6-tri-chloro-1,3,5-triazine (2.08 g, 11.3 mmol) and copper(I) iodide (70.5 mg, 0.37 mmol) were dissolved in 12 mL of dehydrated THF under an Ar atmosphere and the solution was cooled to −10 °C. After 2 h stirring, 13.6 mL (27.1 mmol) of tert-butylmagnesium chloride solution 2.0 M in THF was slowly dropwised to the stirred solution. After stirring at 0 °C for 2 h and at r.t. for 12 h, the reaction mixture was quenched with 50 ml of 2.4 M HCl aqueous solution. The resulting mixture layer was extracted with 50 mL of ethyl acetate three times. The crude mixture was concentrated under reduced pressure and then purified by silica gel column chromatography using hexane/dichloromethane = 9/1 as eluent. 2.32 g (10.2 mmol) of 2,4-di-tert-butyl-6-chloro-1,3,5-triazine (Supplementary Scheme 1) was obtained in 90% yield.

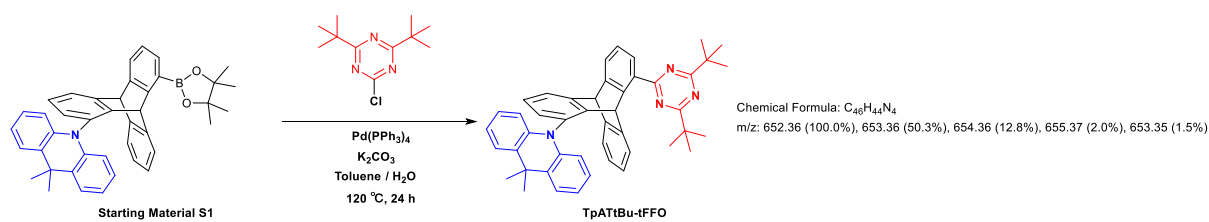

**Supplementary Scheme 2.** Synthesis of TpATtBu-tFFO.

A 45-mL portion of deoxidized toluene and 4.5 mL of 2 M aqueous potassium carbonate (12.0 mmol) were added to a 100-mL round bottom Schlenk flask containing Starting Material S1 (0.42 g, 0.75 mmol), 2,4-di-tert-butyl-6-chloro-1,3,5-triazine (0.29 g, 1.27 mmol), and Pd(PPh$_3$)$_4$ (84 mg, 0.075 mmol). After three freeze-pump-thaw cycles under an Ar atmosphere, the mixture was stirred at 120 °C for 24 h. After the reaction mixture was cooled to ambient temperature, 40 mL of distilled water was added, and the organic layer was extracted with 50 mL of ethyl acetate three times. The combined organics were dried over sodium sulfate, concentrated under reduced pressure and then purified by column chromatography using hexane/dichloromethane = 4/1. 0.366 g (0.56 mmol) of TpATtBu-tFFO was obtained in 74% yield.

$^1$H NMR (400 MHz, CDCl$_3$, δ): 8.11 (dd, J = 8.0, 1.1 Hz, 1H), 7.62-7.60 (m, 1H), 7.53 (d, J = 7.2 Hz, 1H), 7.49 (d, J = 7.0 Hz, 1H), 7.43 (dd, J = 7.8, 1.4 Hz, 1H), 7.32 (d, J = 6.4 Hz, 1H), 7.23-7.15 (m, 3H), 7.13 (s, 1H), 7.10-7.06 (m, 1H), 7.03 (td, J = 7.3, 1.3 Hz, 1H), 6.98-6.94 (m, 1H), 6.86-6.81 (m, 2H), 6.55-6.51 (m, 1H), 6.38-6.34 (m, 1H), 5.84 (dd, J = 8.1, 1.0 Hz, 1H), 5.72 (dd, J = 8.1, 1.0 Hz, 1H), 5.66 (s, 1H), 1.89 (s, 3H), 1.26 (s, 3H), 1.17 (s, 18H). $^{13}$C NMR (101 MHz, CDCl$_3$, δ): 184.1, 171.3, 148.1, 147.2, 146.4, 145.8, 144.6, 143.8, 140.9, 140.5, 136.1, 133.4, 130.6, 129.9, 128.5, 128.0, 126.7, 126.5, 126.2, 126.1, 125.7, 125.2, 125.2, 125.2, 123.7, 123.5, 123.4, 123.4, 120.8, 120.3, 114.8, 113.3, 77.3, 77.0, 76.7, 54.8, 45.6, 39.2, 36.1, 33.0, 28.8, 23.6.

APCI-MS (m/z): [M+H]$^+$ calcd. for C$_{46}$H$_{45}$N$_4$, 653.3639; found, 653.3636.

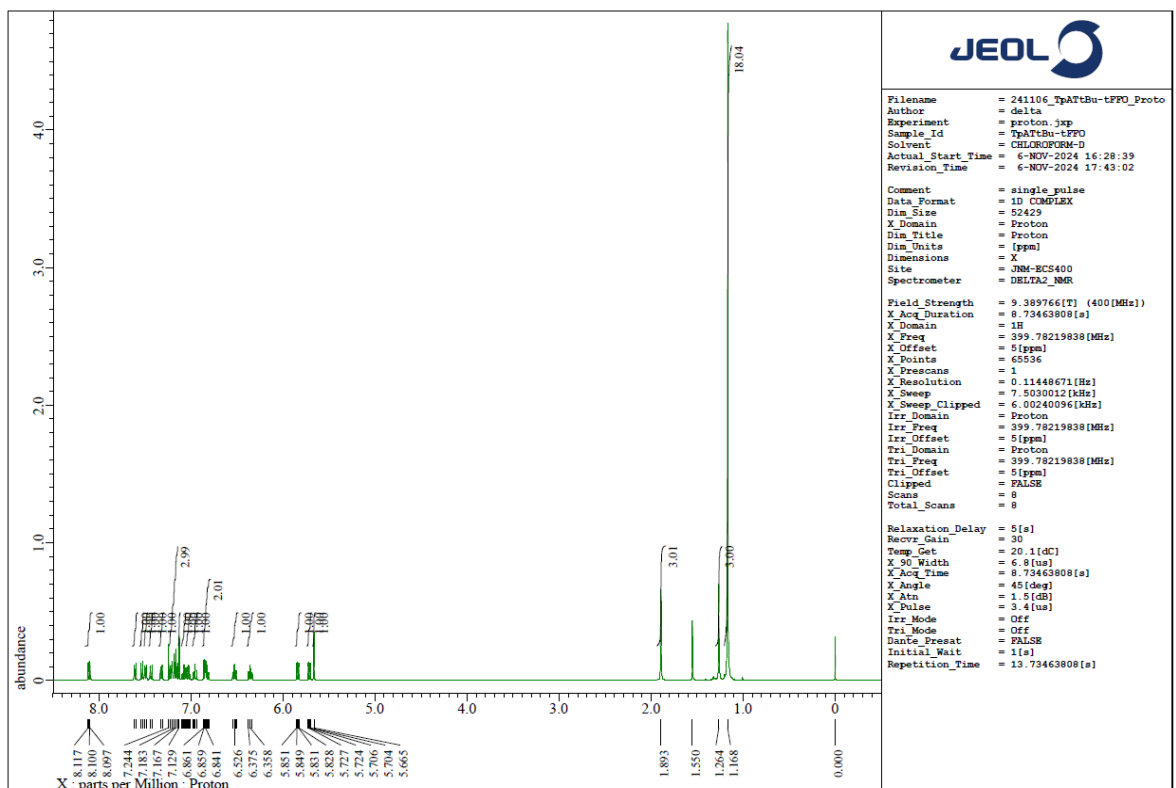

**Supplementary Fig. 7.** $^1$H NMR spectrum of TpATtBu-tFFO.

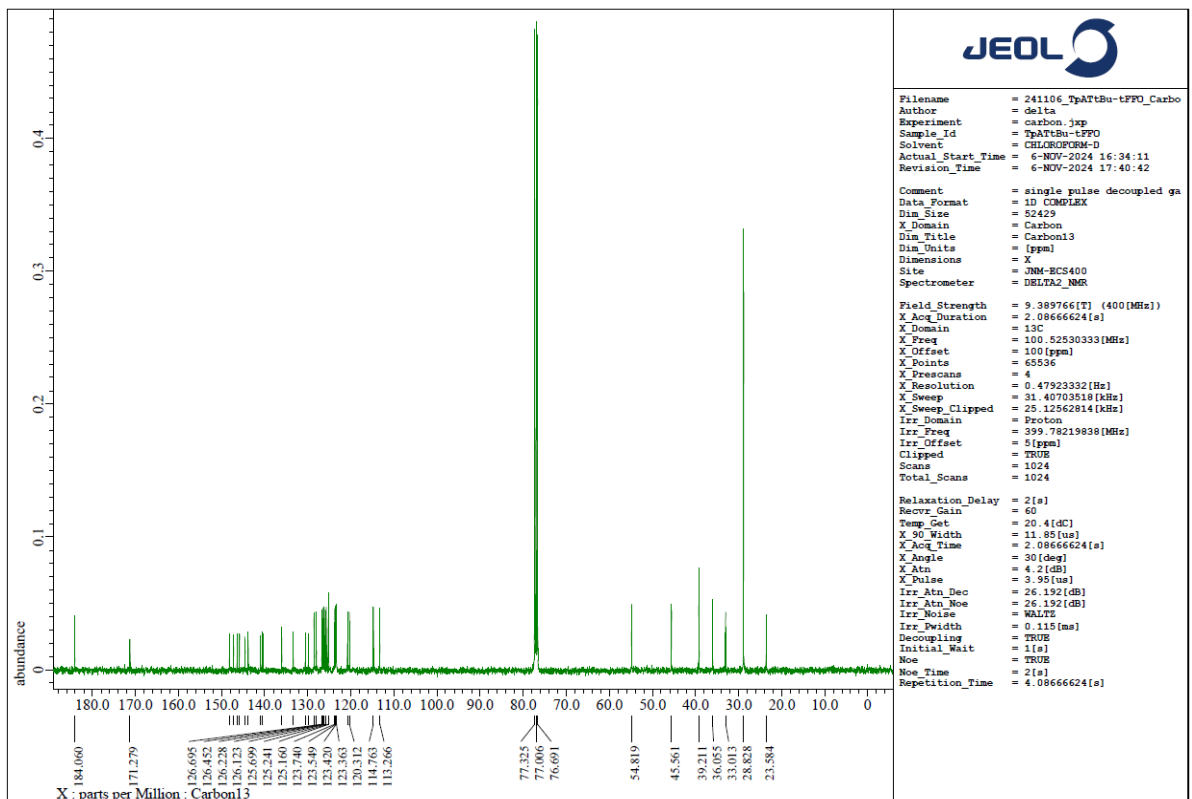

**Supplementary Fig. 8.** $^{13}C\{^{1}H\}$ NMR spectrum of TpATtBu-tFFO.

## References

1. Y. Wada, H. Nakagawa, S. Matsumoto, Y. Wakisaka and H. Kaji, *Nat. Photon.*, 2020, **14**, 643-649.